\documentclass[12pt,preprintnumbers,eqsecnum,nofootinbib]{revtex4}
\usepackage{amsfonts}
\usepackage{mathrsfs}
\usepackage{amssymb}
\usepackage{amsmath}
\usepackage{graphicx}
\usepackage{graphicx}
\usepackage{epstopdf}
\usepackage{pstricks}
\usepackage{slashed}
\usepackage{mathbbol}
\usepackage{color}
\usepackage{refstyle}

\begin{document}
\preprint{OSU-HEP-16-06}
\title{Neutrino Mass Generation and 750 GeV Diphoton excess via photon-photon fusion at the Large Hadron Collider}

\author{Kirtiman Ghosh$^{\blacktriangle}$ \footnote{Email: kirti.gh@gmail.com} , Sudip Jana$^{\bigstar}$ \footnote{Email: sudip.jana@okstate.edu} and  S. Nandi$^{\bigstar}$ \footnote{Email: s.nandi@okstate.edu}
}

\affiliation{$^\bigstar$Department of Physics and Oklahoma Center for High Energy Physics,
Oklahoma State University, Stillwater, OK 74078-3072, USA \\
$^\blacktriangle$Department of Physics and Astrophysics, University of Delhi, Delhi-110007.\\.
}

\date{\today}

\begin{abstract}

\section*{Abstract}
We propose a new model for  explaining the diphoton excess observed at the LHC. The uniqueness of our model is that this excess is being produced by photon-photon induced sub-process. The model includes an isospin $3/2$ scalar  multiplet (having triply, doubly and singly charged scalars) at the electroweak  (EW) scale  to explain the observed neutrino masses via the effective dimension $7$ operator. The model also include a singlet scalar which is the di-photon resonance. The loop induced coupling of the singlet scalar with a pair of photons is enhanced due to contributions from multi-charged EW-scale scalars (component of isospin $3/2$ multiplet) in the loop. We examine the production of the singlet scalar at the LHC via photon-photon fusion sub-process and its subsequent decay to di-photon. We found that a large part of parameter space of the present model can explain the observed excess of di-photon events at the LHC. We also study the other decay modes of the singlet scalar and the ensuing predictions can be tested in the upcoming runs at the LHC.

\end{abstract}


\maketitle

\section{Introduction}
In the 8 TeV runs, and early 13 TeV runs at the Large Hadron Collider, both the ATLAS and CMS Collaborations has reported a significant excess of di-photon events with an invariant mass of about 
750 GeV \cite{at, CMS:2015dxe, Khachatryan:2015qba, Aad:2015mna}. This is probably the first substantial hint at the LHC for any new physics beyond the Standard Model (SM). A large number of papers have appeared in the past seven months and a comprehensive list of citations can be found in \cite {Strumia:2016wys}. Both the experimental status and the various theoretical approaches have also been nicely reviewed in \cite{Strumia:2016wys}. In most of the theoretical explanations, the diphoton excess is being caused by a scalar resonance with a mass of 750 GeV. This is being produced by strong interaction (by the gluon gluon sub-process) via the triangle loop of  heavy vector-like quarks, and then decaying to two photons via the same loop. The main problem of this mechanism is that one has to assume many vector-like quarks, otherwise the coupling of the vector-like quark with the scalar singlet becomes way non-perturbative. Another interesting approach that has been discussed is the photon initiated sub-process instead of the popular gluon initiated sub-process to produce this singlet scalar \cite{Csaki:2016raa, Abel:2016pyc, Dalchenko:2016dfa, Kanemura:2015bli}. 

In this work, we discuss di-photon signature of a 750 GeV scalar resonance in the framework of a model, a part of which was originally proposed to generate a tiny neutrino mass with effective dimension seven operator \cite{Babu:2009aq}. 
In the present work, we add an EW singlet scalar, which is the di-photon resonance, to this model. Photon-photon fusion is the dominant production mechanism for this model at the LHC. In addition to the SM particles, the model includes an isospin $3/2$ scalar  ($\Delta_{3/2}$) and two triplet vector-like leptons. The most general scalar potential includes dimension-3 interactions involving the singlet scalar and $\Delta_{3/2}$. Being induced via loops involving multicharged (triply, doubly and singly) isosipn $3/2$ scalars, one expects significant enhancement in $S\gamma \gamma$ coupling. Therefore, it is instructive to study the di-photon signature of the scalar-$S$ in the context of recent LHC reports of di-photon excess at 750 GeV. Being singlet, the scalar-$S$ does not couple to the SM gauge bosons at tree level. Moreover, the Yukawa interactions involving $S$ and the SM fermions are forbidden due to the chiral nature of the SM fermions. Therefore, at the LHC, only production mechanism possible for the scalar-$S$ is the photon-photon fusion where the photons in the initial state arises either as a parton in side the proton or proton being electrically charged, radiated from it. Though parton density of photon is usually small and photon radiation probability is suppressed by the form factors, the large loop induced $S\gamma \gamma$ coupling could give rise to significant production cross-section at the LHC and hence, di-photon being the dominant decay channel, $S$ could explain the LHC observed excess of di-photon events at 750 GeV invariant mass. The model has several interesting predictions, such as other observable modes with cross section at the few $fb$ level are $h h$, $Z Z$, and $W W$, while the contribution to di-jet and $Z \gamma$ modes are very small. We also emphasize the model can explain both the 750 GeV di-photon rate, as well as the tiny neutrino masses.

\section{Model and Formalism}

Our model is based on the SM symmetry group $SU(3)_{C}\times SU(2)_{L}\times U(1)_{Y}$ . In the fermion sector, in addition to the SM fermions, we add two vector-like $SU(2)$ triplet leptons, $\Sigma $ and $\bar{\Sigma}$. In the scalar sector, in addition to the usual SM Higgs doublet, $H$,  we introduce an isospin $3/2$ scalar, $\Delta_{3/2}$, and and an electroweak (EW) singlet, $S$. The particle contents along with their quantum numbers are shown in the Table~\ref{Table1} below. \\

\begin{table}[htb]
\begin{center}
\begin{tabular}{|c|c|c|}
\hline 
&$SU(3)_{C} \times SU(2)_{L} \times U(1)_{Y}$\\\hline
\small Matter &${\begin{pmatrix} u \\ d \end{pmatrix}}_L\sim(3,2,\frac{1}{3}), u_R\sim (3,1,\frac{4}{3}), d_R \sim (3,1,-\frac{2}{3})$ \\
&$ {\begin{pmatrix} \nu_e \\ e \end{pmatrix}}_L\sim (1,2,-1), e_R\sim (1,1,-2), \nu_R\sim (1,1,-2)$ \\&$ {\begin{pmatrix} \Sigma^{++} \\ \Sigma^{+} \\ \Sigma^{0} \end{pmatrix}}\sim (1,3,2)$, $ {\begin{pmatrix} \bar{\Sigma}^{0} \\ \bar{\Sigma}^{-} \\ \bar{\Sigma}^{--} \end{pmatrix}}\sim (1,3,-2)$ \\
\hline
\small Gauge & $G^\mu_{a,a=1-8}, A^\mu_{i, i=1-3}, B^\mu$ \\
\hline
\small Higgs & ${\begin{pmatrix} \phi^{+} \\ \phi^{0} \end{pmatrix}}\sim(1,2,1)$, ${\begin{pmatrix} \Delta^{+++} \\ \Delta^{++} \\ \Delta^{+} \\\Delta^{0}\end{pmatrix}}\sim(1,4,3)$, $ S\sim(1,1,0)$\\
\hline
\end{tabular}
\\
\end{center}
\caption{Matter, gauge and Higgs contents of the model.}
\label{Table1}
\end{table}

The most general renormalizable  scalar potential  consistent with   {scalar spectrum of this} model is given by,
 {
\begin{eqnarray}
V ( H,\Delta,S)&=&\mu_H^{2}H^{\dagger}H + \mu_\Delta^{2}\Delta^{\dagger}\Delta+\mu_S^{2}S^{2}\nonumber\\
&+& \lambda_{6}M_SS^{3}+ \lambda_{10}M_{S}(H^{\dagger}H)S + \lambda_{11}M_{S}(\Delta^{\dagger}\Delta)S\nonumber\\
&+&   \frac{\lambda_{1}}{2}(H^{\dagger}H)^{2} + \frac{\lambda_{2}}{2}(\Delta^{\dagger}\Delta)^{2} +  \lambda_{3}(H^{\dagger}H)(\Delta^{\dagger}\Delta) \nonumber\\
&+& \lambda_{4}(H^{\dagger}\tau_{a}H)(\Delta^{\dagger}T_{a}\Delta)  + \lbrace\lambda_{5}H^{3}\Delta^{\star} + h.c. \rbrace\nonumber\\
&+& \frac{\lambda_{7}}{2}S^{4} + \lambda_{8}(H^{\dagger}H)S^{2} + \lambda_{9}(\Delta^{\dagger}\Delta)S^{2},
\label{scalar_potential}
\end{eqnarray}
}
where  $\tau_{a}$ and $T_{a}$   {are} the generators of $SU(2)$ in the doublet  {and} four-plet representations, respectively, $M_{S}$ is the mass of scalar singlet ($\sim$ 750 GeV)  {and $\lambda$'s are dimensionless parameters}. \\

As was shown in \cite{Babu:2009aq}, even with positive ${\mu_{\Delta}}^2$, due to the $\lambda_5$ term in the potential,  and the fields $\Sigma$ and $\bar\Sigma$  , the neutral component of $\Delta$ acquires an induced VEV  at the tree level, $v_{\Delta} = - \lambda_5 v^3 / M_{\Delta}^2$, where v is the usual EW VEV. This gives rise to effective dimension seven operator $LLHH(H^{\dagger} H)/ M^{3}$, and generate tiny neutrino masses \cite{Babu:2009aq}. The additional singlet S that we have introduced gets no VEV ($v_s = 0$). Hence, we have as in \cite{Babu:2009aq}
\begin{equation}
 M_{\Delta}^{2} = \mu_\Delta^{2} + \lambda_{3}v_{H}^{2} + \frac{3}{4}\lambda_{4}v_{H}^{2},
 \end{equation}
where $\mu_{\Delta}$ is the mass of the neutral member in $\Delta$. The mass splittings between the members of $\Delta$ are given by
\begin{equation}
M_{i}^{2} = M_{\Delta}^{2} - q_{i}\frac{\lambda_{4}}{4}v_{H}^{2},
\label{spectrum}
\end{equation}
where $q_{i}$ is the (non-negative) electric charge of the respective field. The mass
splittings are equally spaced and there are two possible mass orderings. For $\lambda_{4}$ positive,
we have the ordering $M_{\Delta^{+++}} < M_{\Delta^{++}} < M_{\Delta^{+}} < M_{\Delta^{0}}$ and for $\lambda_{4}$ negative, we have the ordering $M_{\Delta^{+++}} > M_{\Delta^{++}} > M_{\Delta^{+}} > M_{\Delta^{0}}$.
Due to cubic coupling term $\lambda_{10}M_{S}$, there is a mixing ($\beta$) between standard model Higgs (H) and scalar singlet (S). Considering a small VEV of $\Delta$ ($\sim$MeV), we can express ($\beta$) in term of  $\lambda_{10}$ as :
\begin{equation}
\beta = \frac{1}{2} \arctan{(\dfrac{2 M_{S}\lambda_{10} v_{H}}{\sqrt{( M_{S}^{2}-M_{H}^{2})^{2} - 4(M_{S}\lambda_{10})^{2}v_{H}^{2}}})},
\end{equation}
where $M_{H} = 125 GeV$ and $M_{S} = 750 GeV$.

\begin{figure}[h]
\centering
\includegraphics[scale=0.27]{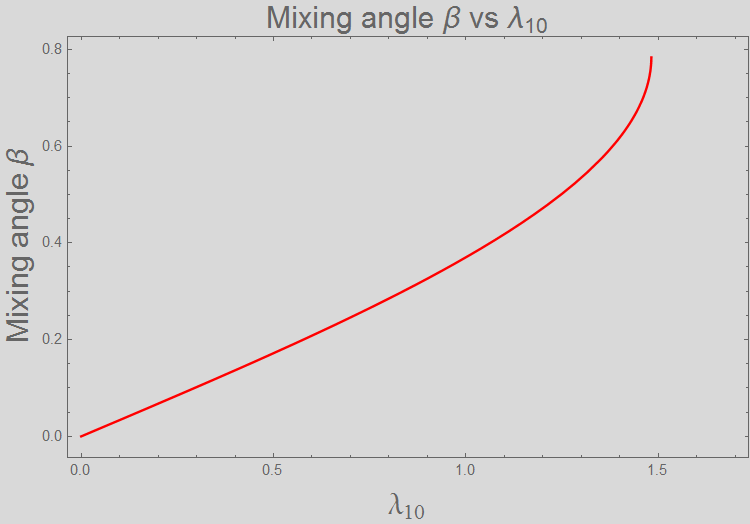}
\caption{Mixing angle (in radian) variation as a function of $\lambda_{10}$}
\label{fig:plot1}
\end{figure}

\section{Diphoton Signal Analysis and Other Collider Implications}

\begin{figure}[h]
\centering
\includegraphics[scale=0.27]{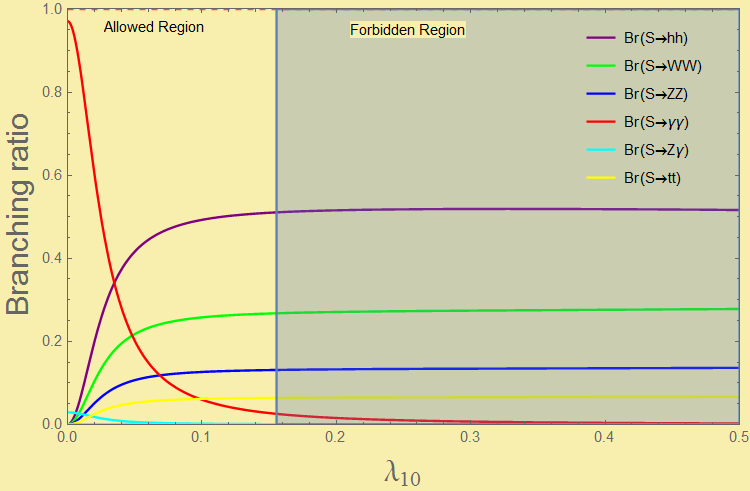}
\caption{Branching ratios of $S$ as a function of $\lambda_{10}$. Blue shaded region is excluded by experimental bounds on partial decay widths for various decay modes.}
\label{fig:plot2}
\end{figure}

In this section, we will discuss the production and decay of the singlet scalar-$S$ in the context of the LHC experiment. Being singlet under the SM gauge group, $S$ does not couple to the gluons or EW-gauge bosons at tree level. Gauge symmetry also forbids its Yukawa couplings with SM quarks and leptons. The Yukawa couplings involving vector-like fermions ($\Sigma$ or $\bar\Sigma$) and singlet-$S$ are allowed, however, irrelevant in the context of collider phenomenology because we assume, for simplicity,  $\Sigma~(\bar \Sigma)$ masses to be large ($\sim$ 1.5 TeV), or larger . The only relevant interactions, in the context of collider phenomenology of $S$, are the dimension 3 terms in the scalar potential involving pair of SM Higgs doublet and pair of TeV scale isospin $3/2$ scalar, $\Delta_{3/2}$ (see Eq.~\ref{scalar_potential}).  The former gives rise to tree-level decay of $S$ into Higgs pairs whereas, the later induce interactions with a pair of photons, $Z$-bosons, $W$-bosons, $Z\gamma$, hh and Zh-pairs via loops involving components of $\Delta_{3/2}$ and hence, allows $S$ to decay into $\gamma \gamma$, $Z\gamma$, $ZZ$, $W^\pm W^\mp$, hh and Zh pairs. However, only loop induced  $\gamma \gamma$ and $Z\gamma$ decays are enhanced due to large charge of isospin $3/2$ scalars. The contribution to the $W^\pm W^\mp$ and $ZZ$ decay modes of $S$ also arises from its mixing with the SM Higgs boson which has a tree level gauge coupling with $W^\pm W^\mp$ and $ZZ$. As a consequence of this mixing, $S$ is also allowed to decay into a pair of top quarks, bottom quarks, tau leptons and even into a pair of gluons. As discussed in the previous section, the mixing depends on the parameter $\lambda_{10}$. For small $\lambda_{10}$ and hence, for small mixing, the mixing induced decays are highly suppressed. Therefore, for small $\lambda_{10}$, the dominant decay modes are tree-level decay to $hh$ and loop induced decay to $\gamma \gamma$. It is important to note that the loop induced coupling of $S$ with a pair of photon is proportional to the square of the electric charge of the loop particle and hence, enhanced by more than two orders of magnitude in the framework of this model due to triply charged scalar $\Delta^{\pm\pm\pm}$. The partial decay width of $S$ into a pair of photons is given by \cite{Djouadi:2005gj,Carena:2012xa}
\begin{equation}
\Gamma(S\rightarrow\gamma\gamma) = \dfrac{\alpha^{2}M_{S}^{3}}{1024\pi^{3}}\left| \sum_{i=1}^{3} Q_{\Delta_{i}}^{2} \dfrac{M_{S}\lambda_{11}}{M_{\Delta_{i}}^{2}}F_{0}(x_{\Delta_{i}}) \right|^{2},
\end{equation}
where $Q_{\Delta_{i}}$ and $M_{\Delta_i}$ stands for the electric charge and mass of  the three charged components $(\Delta^{+++},~\Delta^{++}~{\rm and}~\Delta^{+})$ of isospin $3/2$ scalar multiplet. The loop function is given by \cite{Djouadi:2005gj,Carena:2012xa} $F_{0}(x_{\Delta}) = \tau [1-\tau f(\tau)],$ where $\tau = 4M_{\Delta}^{2}/M_{S}^{2}$ and $f(\tau) = [\sin^{-1}(\sqrt{\tau^{-1}})]^{2}$ for $\tau \geq 1$. Splitting between the components of $\Delta_{3/2}$ depends on $\lambda_{4}$ in Eq.~\ref{scalar_potential} which is constrained to be small from the experimental limits on $\rho$-parameter. Therefore, for small $\lambda_4$, the spectrum of isospin $3/2$ scalars (see Eq.~\ref{spectrum}) is quasi-degenerate. In our analysis, we consider $M_{\Delta}>M_S/2$ in order to prohibit the tree-level decay of $S$ into a pair of $\Delta$'s. On the other hand, in order to maximize the loop function we choose $M_\Delta \sim M_S/2 = 376$ GeV. For a fixed $M_\Delta$, the total decay width and hence, the branching ratios of $S$ depends on $\lambda_{10}$
\footnote{$\lambda_{10}$ determines the mixing between $S$ and $h$ and hence, the mixing induced decay widths like $\Gamma(t\bar t)$, $\Gamma(b\bar b)$, $\Gamma(\tau \bar \tau)$, $\Gamma(gg)$, $\Gamma(W^\pm W^\mp)$ and $\Gamma(ZZ)$.}  and $\lambda_{11}$\footnote{The di-photon and $Z\gamma$ decay width is proportional to $\lambda_{11}^2$. $\Gamma(W^\pm W^\mp)$ and $\Gamma(ZZ)$ decay modes also get loop induced contributions proportional to $\lambda_{11}^2$.}. In Fig.~\ref{fig:plot2}, we have presented the branching ratios of $S$ in to different decay channels as a function of $\lambda_{10}$ for fixed $\lambda_{11}=2$. For $\lambda_{10}=0$, there is no mixing and thus, the only possible decay modes are $\gamma\gamma$ and $Z\gamma$. $\Gamma(\gamma\gamma)$ is enhanced by $Q^4$, where $Q$ is the charge of the loop scalar, whereas, $\Gamma(Z\gamma)$ is enhanced by $Q^2$ only. As a result, for $\lambda_{10}=0$, branching ratio to $\gamma \gamma$ is almost 100\%. However, the situation changes drastically as we increase $\lambda_{10}$ and hence, increase the tree-level decay to $hh$ and mixing induced decays into $ZZ,~W^\pm W^\mp$ and $t\bar t$. The bound (also shown in Fig.~\ref{fig:plot2}) on $\lambda_{10}$ comes from the experimental limits on $\Gamma(hh)/\Gamma(\gamma\gamma)< 20$ \cite{Franceschini:2015kwy, Liao:2015tow} for a 750 GeV resonance consistent with the LHC di-photon excess.

\begin{figure}[h]
\centering
\includegraphics[scale=0.3]{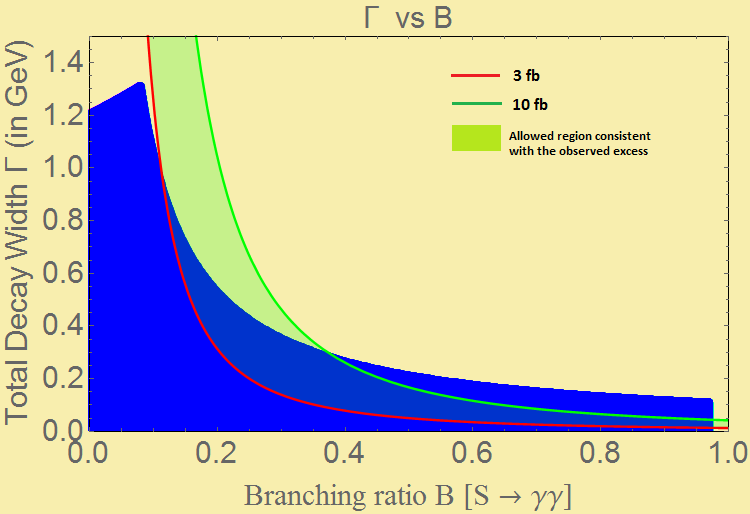}
\caption{Allowed region in ${\rm Br}(S\to \gamma\gamma)$--$\Gamma_{\rm TOT}$ plane consistent with the observed excess ($\sim$ 3--10 fb) of di-photon events at the LHC with $\sqrt{S}$= 13 TeV is indicated by the region specified in between 3 fb and 10 fb lines. The blue shaded region is our model prediction for $\Gamma_{\rm TOT}$ and ${\rm Br}(S\to \gamma\gamma)$ as we scan the parameter space over $\lambda_{10}$ (between 0--0.15) and $\lambda_{11}$ (between 0--3.5).}
\label{fig:plot3}
\end{figure}

\begin{figure}[h]
\centering
\includegraphics[scale=0.3]{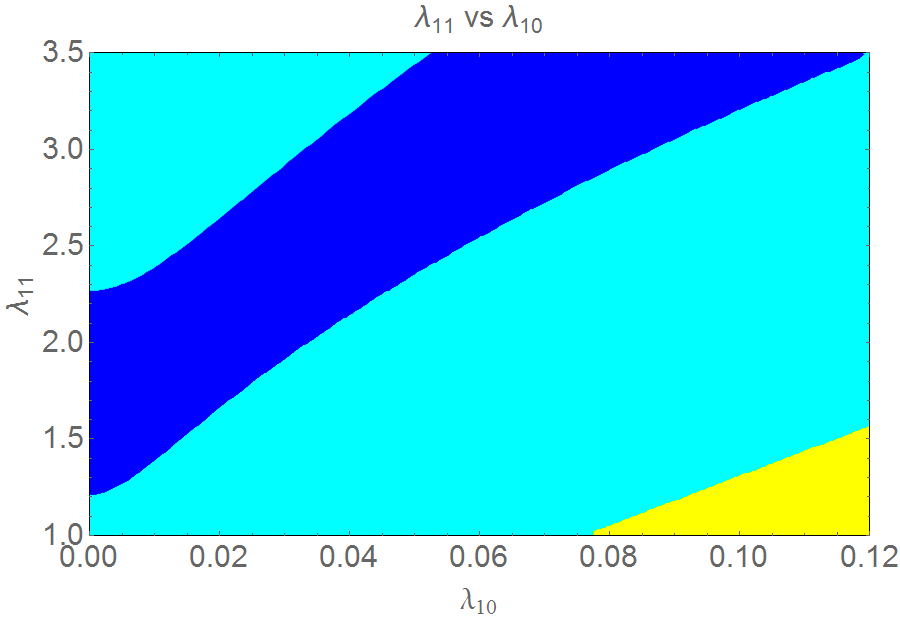}
\caption{The part of $\lambda_{10}$--$\lambda_{11}$ plane consistent with LHC 13 TeV excess (3--10 fb) of di-photon events is indicated by blue shaded region. The yellow-shaded region is excluded from experimental bounds on various other decay modes of $S$.}
\label{fig:plot4}
\end{figure}

The scalar-$S$ does not have any loop induced coupling to a pair of gluons. Coupling to a pair of gluons arises due to mixing with the SM Higgs and hence, suppressed for small $\lambda_{10}$. However, as discussed in the previous paragraph, the loop induced coupling of $S$ with a pair of photons is enhanced due to multi-charged isospin $3/2$ scalars in the loop. Therefore, photoproduction {\em i.e.,} production via photon-photon fusion, at the LHC is the dominant production mechanism for $S$. The photoproduction cross-section of $S$ can be estimated from its decay width into a pair of photons and photon distribution in proton.  The di-photon signal cross-section is given by \cite{Kanemura:2015bli, Abel:2016pyc, Csaki:2016raa},
\begin{equation}
\sigma(pp \to S\to \gamma\gamma) = \sigma_0 \left(\frac{\Gamma_{\rm TOT}}{{\rm GeV}}\right){\rm Br}^2(S\to \gamma \gamma),
\label{cross}
\end{equation} 
where, $\Gamma_{\rm TOT}$ is the total decay width of resonance $S$ and $\sigma_0$ gets contribution from fully inelastic ($\sim 63\%$), partially inelastic ($\sim 33\%$) and elastic ($\sim 4\%$) proton-proton collision and estimated to be about 240 (122) fb \cite{Abel:2016pyc,Csaki:2016raa} at the LHC with $\sqrt s=13~(8)$ TeV. Eq.~\ref{cross} shows that large total decay width of resonance $R$ and/or sizable branching ratio to di-photon could explain the size (3--10 fb) of the observed di-photon excess via photoproduction of $S$ at the LHC. In Fig.~\ref{fig:plot3}, we have presented the region of ${\rm Br}(S\to \gamma\gamma)$--$\Gamma_{\rm TOT}$ plane consistent with 3--10 fb excess of di-photon events at 750 GeV observed at the LHC. We also scan the parameter space of our model over $\lambda_{10}$ (between 0--0.15) and $\lambda_{11}$ (between 0--3.5) and presented the resulting total decay widths and di-photon branching ratios in Fig.~\ref{fig:plot3}. There is significant overlap between allowed region (green shaded region) and our model prediction (blue shaded region) in ${\rm Br}(S\to \gamma\gamma)$--$\Gamma_{\rm TOT}$ plane which indicates that there is a finite part of parameter space in our model consistent with the di-photon excess at 13 TeV LHC. In order to pin down the part of $\lambda_{10}$--$\lambda_{11}$ plane consistent with the observed $\gamma \gamma$ excess (3--10 fb), we have presented Fig.~\ref{fig:plot4}.  The blue shaded region in $\lambda_{10}$--$\lambda_{11}$ plane (see Fig.~\ref{fig:plot4}) gives rise to 3--10 fb excess of di-photon events with an invariant mass of $\sim$ 750 GeV at the LHC with 13 TeV center-of-mass energy. In Fig.~\ref{fig:plot4}, we have also shown the part of $\lambda_{10}$--$\lambda_{11}$ plane excluded from the experimental limits \cite{Strumia:2016wys, Aad:2014fha, Aad:2015agg, Aad:2015kna, Aad:2015fna} on other decay modes of $S$. In Table.~\ref{table1}, we have presented our model predictions for cross-sections for different decay modes of $S$ for three different benchmark points (BP's). The last column of Table~\ref{table1} corresponds to the experimental limits on the respective cross-sections. 

\begin{table}[h]
  \renewcommand{\arraystretch}{0.001}
  \begin{tabular}{|p{2.5 cm}|c|c|c|c|}
    \hline
    \textbf{ Final states} & \multicolumn{3}{c|}{\textbf{Model prediction for $\sigma$ at $\sqrt{S}$ = 13 TeV [fb]}} & \textbf{Observed experimental limit}\\
    \cline{2-4}
    & \textbf{BP1} & \textbf{BP2} & \textbf{BP3} & \textbf{[fb]}\\
     & \textbf{$\lambda_{10} = 0$} & \textbf{$\lambda_{10} = 0.05$} & \textbf{$\lambda_{10} = 0.11$} & \textbf{}\\
      & \textbf{$\lambda_{11} = 2$} & \textbf{$\lambda_{11} = 3$} & \textbf{$\lambda_{11}= 3.5$} & \textbf{}\\
    \hline
     $ \gamma\gamma$ & 7.93 & 6.61 & 3.45 & 6.26$\pm$3.32 \\
     ZZ & $\sim 0$ & 1.59 & 2.89 & $<19$\\
     Z$\gamma$ & 0.237 & 0.198 & 0.103 & $<28$\\
     hh & $\sim 0$ & 6.074 & 11.26 & $<120$ \\
     $W^{+}W^{-}$ & $\sim 0$ & 3.18 & 5.89 & $<37$ \\
     $t\bar{t}$ & 0 & 0.759 & 1.409 & $<700$\\
     gg & 0 & 1.55$\times10^{-4}$ & 2.88$\times10^{-4}$ & $<2200$ \\
      \hline
  \end{tabular}
  \caption{Bounds on cross-section for various final states.}
\label{table1}
\end{table}

\section{Conclusions}
We have presented new model for the recent di-photon excess at the LHC. The uniqueness of the model is that it can explain not only the diphoton excess, but also generate tiny neutrino masses via the see-saw mechanism at the TeV scale. Another uniqueness of the model is that the diphoton excess is generated by the two photon induced subprocess. The model has a SM singlet scalar which is the diphoton resonance, and a weak isospin $3/2$ scalar with multiply charged particles circulating 
in the loop for the production as well as the decay. The model satisfies all the constraints associated with the diphoton events. It also predicts the other observable final states to be $h h$, $Z Z$, $W W$ and almost no dijet resonance and very tiny cross sections for $Z \gamma$. All these predictions can be tested in the upcoming runs with higher luminosity at the LHC.


\begin{acknowledgments}
 The work of SN was 
in part supported by US Department of Energy Grant rant Numbers  DE-SC 0010108 and DE-SC0016013. The work of KG was supported by an Inspire Faculty Fellowship from the Government of India, at the University of Delhi.

\end{acknowledgments}

\end{document}